# Discovery of a novel 112-type iron-pnictide and La-doping induced superconductivity in $Eu_{1-x}La_xFeAs_2$ (x = 0 ~ 0.15)

Jia Yu[1], Tong Liu[1], Bo-Jin Pan[1], Bin-Bin Ruan[1], Xiao-Chuan Wang[1], Qing-Ge Mu[1], Kang Zhao[1], Gen-Fu Chen[1], Zhi-An Ren[1, 2, 3] *

[1] Institute of Physics and Beijing National Laboratory for Condensed Matter Physics, Chinese Academy of Sciences, Beijing 100190, China

[2] Collaborative Innovation Center of Quantum Matter, Beijing 100190, China

[3] School of Physical Sciences, University of Chinese Academy of Sciences

* Email: renzhian@iphy.ac.cn

**Abstract:**

We report the discovery and characterization of a novel 112-type iron pnictide $EuFeAs_2$, with La-doping induced superconductivity in a series of $Eu_{1-x}La_xFeAs_2$. The polycrystalline samples were synthesized through solid state reaction method only within a very narrow temperature window around 1073 K. Small single crystals were also grown from a flux method with the size about 100 μm. The crystal structure was identified by single crystal X-ray diffraction analysis as a monoclinic structure with space group of $P2_1/m$. From resistivity and magnetic susceptibility measurements, we found that the parent compound $EuFeAs_2$ shows a $Fe^{2+}$ related antiferromagnetic/structural phase transition near 110 K and a $Eu^{2+}$ related antiferromagnetic phase transition near 40 K. La doping suppressed the both phase transitions and induced superconducting transition with a $T_c$ ~ 11 K for $Eu_{0.85}La_{0.15}FeAs_2$.

## 1. Introduction

Since the first report of high-$T_c$ superconductivity in LaFeAs(O,F) at 26 K in 2008, hundreds of new layered iron pnictide superconductors have been discovered, with all of them formed by the typical $Fe_2As_2$ layers stacking with various types of blocking layers [1-10]. Among them, the 1111-type SmFeAs(O,F) still holds the highest $T_c$ around 55 K in bulk materials after 8 years [11]. Meanwhile, superconducting evidences around 100 K was reported in single unit cell film of FeSe on $SrTiO_3$ substrate [12]. Besides, some other layered transition metal pnictide superconductors were also discovered with much lower superconducting $T_c$ [13-15]. Therefore, it is still challenging to design and explore new iron-based superconductors with higher $T_c$ or novel structures for practical applications and physical research.

In all these iron pnictides, the newly reported 112-type $Ca_{1-x}RE_xFeAs_2$ (RE = La-Gd) superconductors are quite unique in crystal structure, which adopts a monoclinic space group with alternatively stacked $Fe_2As_2$ layers and Ca-$As_2$-Ca blocks, where the $As_2$ layers are made of unique zigzag As-As chains, with monovalent $As^-$ anions comparing with the trivalent $As^{3-}$ in $Fe_2As_2$ layer [9, 16-19]. Due to the special staggered intercalation layered crystal structure, recently $CaFeAs_2$ was also predicted to be an ideal candidate that integrates topological quantum spin Hall and superconductivity for the exploration of Majorana fermions [20, 21]. Unfortunately, the chemical phase of undoped $CaFeAs_2$ has never been obtained yet, and the only synthesized stable phases are the rare earth doped ones. The doping of rare earth metals introduce electrons and bring superconductivity in these 112-type compounds, while more doping of electrons may suppress the superconductivity and enhance the antiferromagnetism in the $Fe_2As_2$ layer unexpectedly [22, 23].

In this report, we have systematically studied the reacting conditions for the Eu-Fe-As ternary phases, and successfully synthesized the 112-type parent compound $EuFeAs_2$ for the first time. Through La-doping, superconductivity was induced in the $Eu_{1-x}La_xFeAs_2$ compounds with the $T_c$ up to 11 K.

## 2. Experimental details

The Polycrystalline samples for $Eu_{1-x}La_xFeAs_2$ ($x$ = 0, 0.05, 0.1, 0.15) were synthesized by the solid state reaction method using EuAs, LaAs, and FeAs as precursors. The precursor materials were reacted by stoichiometric metal with Arsenic powder, which were mixed thoroughly and pressed into pellets, placed in alumina crucibles and sealed into Argon filled quartz tubes. The tubes were slowly heated to 1123 K and held for two days in a furnace. The powder of precursors were mixed in stoichiometric ratio of $Eu_{1-x}La_xFeAs_2$, and pressed into pellets. The pellets were placed in alumina crucibles and sealed into Argon filled quartz tubes. The tubes were slowly heated to 1073 K and held for one week. Single crystals of $Eu_{1-x}La_xFeAs_2$ were also attempted to grow by flux methods, and the grown crystals are with the size about 100 μm. All preparing manipulations were carried out in a glove box protected with high-purity argon gas.

Single crystal X-ray diffraction (SXRD) and powder X-ray diffraction (PXRD) were used for the identification of crystal structure and chemical phases. The sample resistivity was measured by the standard four-probe method using a Quantum Design physical property measurement system (PPMS), and the magnetic susceptibility was measured by a Quantum Design magnetic property measurement system (MPMS).

## 3. Results and discussions

The crystal structure for the $Eu_{1-x}La_xFeAs_2$ samples was determined by SXRD and PXRD analysis. Since the crystal growth is difficult, only one piece of well-shaped crystal with nominal composition of $Eu_{0.9}La_{0.1}FeAs_2$ was used for SXRD experiment, and the optical image of the crystal is shown in Fig. 1(a). The atomic composition for this crystal was analyzed by energy-dispersive X-ray spectroscopy (EDX) method, with the atomic ratio of Eu : La : Fe : As = 20.2 : 2.3 : 24.3 : 53.2. This result confirmed the 112-type composition, and the La-doping level is about 10.2%. Therefore, the nominal formula was used hereinafter. The crystallographic data obtained from the SXRD analysis, the refined conditions, as well as the refined atomic coordinates and equivalent isotropic atomic displacement parameters were summarized in Table I. It confirms the chemical formula and reveals that the $Eu_{0.9}La_{0.1}FeAs_2$ crystal crystallizes in a monoclinic structure with space group $P2_1/m$ (No. 11), in accordance with the results of previously reported $(Ca, Pr)FeAs_2$ [16]. The crystal structure was depicted in Fig. 1(b).

All polycrystalline $Eu_{1-x}La_xFeAs_2$ samples were characterized by PXRD for structure analysis and phase identification. The PXRD patterns are presented in Fig. 1(c). Small amount of $Eu_2O_3$ and $FeAs_2$ impurities can be observed, and more impurities appear when the doping level x increases. All the PXRD patterns were well indexed with the space group $P2_1/m$. The lattice parameters for the parent compound of $EuFeAs_2$ are refined to be $a$ = 3.980(0) Å, $b$ = 3.900(6) Å, $c$ = 10.643(9) Å, and $\beta$ = 90.035(1) °. For different La-doping levels, the lattice parameters have no obvious change due to the close radiuses of $Eu^{2+}$ and $La^{3+}$ ions. The La-doping levels were confirmed by EDX analysis, and we found the real La-doping levels are all close to the nominal ones in the polycrystalline samples.

The normalized resistivity data vs. temperature for all polycrystalline $Eu_{1-x}La_xFeAs_2$ samples were plotted in Fig. 2. All samples show metallic behavior below 300 K. For the undoped parent compound $EuFeAs_2$, two distinct anomalies appear around $T_N$ ~ 110 K and $T_N$ ~ 40 K respectively. The anomaly at 110 K is due to the $Fe^{2+}$ related antiferromagnetic/structural phase transition, as in other parent compounds of Fe-based superconductors [11, 24-28], as well as in $Ca_{1-x}RE_xFeAs_2$ [23, 29, 30]. The anomaly at 40 K indicates the antiferromagnetic (AFM) phase transition of $Eu^{2+}$ ions, as in $EuFe_2As_2$, $EuCu_2As_2$, $EuPd_2Sb_2$, *etc*. [27, 31, 32]. For the La-doping samples of $Eu_{1-x}La_xFeAs_2$, these two transitions are suppressed and shift to lower temperatures, but not completely eliminated even for $x$ = 0.15. Meanwhile, a superconducting transition appears when $x$ = 0.05, and the onset superconducting $T_c$ reaches 11 K when $x$ = 0.15, which can be seen in the inset of Fig. 2. The coexistence of the superconducting phase with two AFM phases in the $Eu_{1-x}La_xFeAs_2$ system may provide an ideal platform for the investigation of mutual interactions between the ordered electron states.

The temperature dependences of DC magnetic susceptibility were measured between 1.8 K and 300 K by the zero-field-cooling (ZFC) method for all polycrystalline $Eu_{1-x}La_xFeAs_2$ samples. The $Eu^{2+}$ related AFM transitions around $T_N \sim$ 40 K for all samples are clearly revealed with similar transition temperatures indicated by the resistivity measurements, agreeing with the AFM behavior in other Eu-intercalated layered materials [27, 33, 34]. Doping La suppresses this AFM transition to lower temperature until $T_N \sim$ 27 K when $x = 0.15$. The AFM transition around 110 K cannot be clearly observed due to the small magnetic moment of $Fe^{2+}$ comparing with $Eu^{2+}$. A weak superconducting transition appears for La-doping when x = 0.05, and it enhances when x increases, with an onset $T_c$ about 8 K when x = 0.15. While all the polycrystalline $Eu_{1-x}La_xFeAs_2$ superconductors show very weak superconducting shielding volume fraction from the Meissner effect measurements.

To further confirm the La-doping induced superconductivity, the DC magnetic susceptibility for the above-mentioned single crystal $Eu_{0.9}La_{0.1}FeAs_2$ was also measured from 1.8 K to 10 K with both ZFC and FC methods under a 10 Oe magnetic field parallel to the c axis, as seen in Fig. 4. The Meissner effect shows typical type-II superconducting behavior. The onset superconducting $T_c$ is 6 K, and the estimated superconducting shielding volume fraction is 91% at 1.8 K. This indicates that La-doping actually induces bulk superconductivity in these novel $Eu_{1-x}La_xFeAs_2$ compounds.

In summary, a new 112-type iron pnictide $EuFeAs_2$ was discovered, with a monoclinic crystal structure by the space group of $P2_1/m$. The parent compound $EuFeAs_2$ shows two AFM phase transitions originating from the $Fe^{2+}$ layers and $Eu^{2+}$ layers. Upon La-doping, the AFM ordering was gradually suppressed and superconductivity simultaneously appears with a $T_c$ of 11 K in the polycrystalline $Eu_{0.85}La_{0.15}FeAs_2$. Bulk superconductivity was confirmed in the $Eu_{0.9}La_{0.1}FeAs_2$ single crystal from the Meissner effect measurements. Unlike the previously reported meta-stable Ca-112 pnictides whose parent compound $CaFeAs_2$ cannot be obtained, these new Eu-112 pnictides provide a better system to study how the superconductivity evolves from the AFM ground state in the parent compound by electron-doping and possible emergent new physics inside this unique 112-type iron pnictide.

**Acknowledgments**

The authors are grateful for the financial supports from the National Natural Science Foundation of China (No. 11474339), the National Basic Research Program of China (973 Program, No. 2016YFA0300301, 2010CB923000 and 2011CBA00100) and the Youth Innovation Promotion Association of the Chinese Academy of Sciences.

**References:**

1. Kamihara Y, Watanabe T, Hirano M, et al (2008) Iron-based layered superconductor La[O$_{1-x}$F$_x$]FeAs ($x$ = 0.05-0.12) with $T(c)$ = 26 K. Journal of the American Chemical Society 130: 3296-3297
2. Wang XC, Liu QQ, Lv YX, et al (2008) The superconductivity at 18 K in LiFeAs system. Solid State Commun 148: 538-540
3. Rotter M, Tegel M, Johrendt D, et al (2008) Spin-density-wave anomaly at 140 K in the ternary iron arsenideBaFe$_2$As$_2$. Phys Rev B 78: 020503
4. Han F, Zhu XY, Mu G, et al (2008) SrFeAsF as a parent compound for iron pnictide superconductors. Phys Rev B 78: 180503
5. Zhu XY, Han F, Mu G, et al (2009) Sr$_3$Sc$_2$Fe$_2$As$_2$O$_5$as a possible parent compound for FeAs-based superconductors. Phys Rev B 79: 024516
6. Sun YL, Jiang H, Zhai HF, et al (2012) Ba$_2$Ti$_2$Fe$_2$As$_4$O: A new superconductor containing Fe$_2$As$_2$ layers and Ti$_2$O sheets. Journal of the American Chemical Society 134: 12893-12896
7. Kakiya S, Kudo K, Nishikubo Y, et al (2011) Superconductivity at 38 K in Iron-Based Compound with Platinum–Arsenide Layers Ca$_{10}$(Pt$_4$As$_8$)(Fe$_{2-x}$Pt$_x$As$_2$)$_5$. J Solid State Chem 80: 093704
8. Katrych S, Pisoni A, Bosma S, et al (2014) L$_4$Fe$_2$As$_2$Te$_{1-x}$O$_{4-y}$F$_y$ (L = Pr, Sm, Gd)_ A layered oxypnictide superconductor with $T$-$c$ up to 45 K. Phys Rev B 89: 024518
9. Katayama N, Kudo K, Onari S, et al (2013) Superconductivity in Ca$_{1-x}$La$_x$FeAs$_2$: A Novel 112-Type Iron Pnictide with Arsenic Zigzag Bonds. J Phys Soc Jpn 82: 123702
10. Liu Y, Liu YB, Chen Q, et al (2016) A new ferromagnetic superconductor: CsEuFe$_4$As$_4$. Science Bulletin 61: 1213-1220
11. Ren ZA, Lu W, Yang J, et al (2008) Superconductivity at 55K in iron-based F-doped layered quaternary compound Sm[O$_{1-x}$F$_x$]FeAs. Chinese Phys Lett 25: 2215-2216
12. Ge JF, Liu ZL, Liu C, et al (2015) Superconductivity above 100 K in single-layer FeSe films on doped SrTiO3. Nat Mater 14: 285-289
13. Guo Q, Pan BJ, Yu J, et al (2016) Superconductivity at 7.8 K in the ternary LaRu$_2$As$_2$ compound. Sci Bull 61: 921-924
14. Guo Q, Yu J, Ruan B-B, et al (2016) Superconductivity at 3.85 K in BaPd$_2$As$_2$with the ThCr$_2$Si$_2$-type structure. Europhys Lett 113: 17002
15. Xiao-Chuan W, Bin-Bin R, Jia Y, et al (2016) Superconductivity in the ternary iridium-arsenide BaIr$_2$As$_2$. Preprint at https://arxiv.org/abs/1608.08352
16. Yakita H, Ogino H, Okada T, et al (2014) A new layered iron arsenide superconductor: (Ca,Pr)FeAs$_2$. Journal of the American Chemical Society 136: 846-849
17. Wang MT, McDonald R, Mar A (1999) Nonstoichiometric rare-earth copper arsenides RECu$_{1+x}$As$_2$ (RE = La, Ce, Pr). J Solid State Chem 147: 140-145
18. Rutzinger D, Bartsch C, Doerr M, et al (2010) Lattice distortions in layered type arsenides LnTAs$_2$ (Ln=La–Nd, Sm, Gd, Tb; T=Ag, Au): Crystal structures, electronic and magnetic properties. J Solid State Chem 183: 510-520
19. Sala A, Yakita H, Ogino H, et al (2014) Synthesis and physical properties of Ca$_{1-x}$RE$_x$FeAs$_2$ with RE= La–Gd. Appl Phys Express 7: 073102
20. Wu XX, Qin SS, Liang Y, et al (2015) CaFeAs$_2$: A staggered intercalation of quantum spin Hall and high-temperature superconductivity. Phys Rev B 91: 081111
21. Liu ZT, Xing XZ, Li MY, et al (2016) Observation of the anisotropic Dirac cone in the band dispersion of 112-structured iron-based superconductor Ca$_{0.9}$La$_{0.1}$FeAs$_2$. Appl Phys Lett 109: 042602
22. Kawasaki S, Mabuchi T, Maeda S, et al (2015) Doping-enhanced antiferromagnetism inCa$_{1-x}$La$_x$FeAs$_2$. Phys Rev B 92: 180508
23. Jiang S, Liu C, Cao HB, et al (2016) Structural and magnetic phase transitions inCa$_{0.73}$Le$_{0.27}$FeAs$_2$ with electron-overdoped FeAs layers. Phys Rev B 93: 054522
24. Huang Q, Qiu Y, Bao W, et al (2008) Neutron-diffraction measurements of magnetic order and a structural transition in the parent BaFe$_2$As$_2$ compound of FeAs-based high-temperature superconductors. Physical review letters 101: 257003
25. Rotter M, Tegel M, Johrendt D (2008) Superconductivity at 38 K in the iron arsenide (Ba$_{1-x}$K$_x$)Fe$_2$As$_2$. Physical review letters 101: 107006
26. Chen GF, Li Z, Li G, et al (2008) Superconductivity in hole-doped (Sr$_{1-x}$K$_x$)Fe$_2$As$_2$. Chinese Phys Lett 25: 3403-3405

27. Ren Z, Zhu ZW, Jiang S, et al (2008) Antiferromagnetic transition in $EuFe_2As_2$: A possible parent compound for superconductors. Phys Rev B 78: 052501

28. Luo YK, Tao Q, Li YK, et al (2009) Evidence of magnetically driven structural phase transition in RFeAsO (R=La, Sm, Gd, and Tb): A low-temperature x-ray diffraction study. Phys Rev B 80: 224511

29. Yang R, Xu B, Dai YM, et al (2016) Optical study of the antiferromagnetic ordered state in electron-overdoped $Ca_{0.77}Nd_{0.23}FeAs_2$. Phys Rev B 93: 245110

30. Kudo K, Kitahama Y, Fujimura K, et al (2014) Superconducting Transition Temperatures of up to 47 K from Simultaneous Rare-Earth Element and Antimony Doping of 112-Type $CaFeAs_2$. J Phys Soc Jpn 83: 093705

31. Sengupta K, Paulose PL, Sampathkumaran EV, et al (2005) Magnetic behavior of $EuCu_2As_2$: A delicate balance between antiferromagnetic and ferromagnetic order. Phys Rev B 72: 184424

32. Das S, McFadden K, Singh Y, et al (2010) Structural, magnetic, thermal, and electronic transport properties of single-crystal $EuPd_2Sb_2$. Phys Rev B 81: 054425

33. Ballinger J, Wenger LE, Vohra YK, et al (2012) Magnetic properties of single crystal $EuCo_2As_2$. J Appl Phys 111: 07E106

34. Weber F, Cosceev A, Drobnik S, et al (2006) Low-temperature properties and magnetic order of $EuZn_2Sb_2$. Phys Rev B 73: 014427

Table I. Crystal structural parameters and atomic coordinates for $Eu_{0.9}La_{0.1}FeAs_2$ single crystal

| Chemical formula | $Eu_{0.9}La_{0.1}FeAs_2$ |
|---|---|
| Crystal system | Monoclinic |
| Space group | $P2_1/m$ (No. 11) |
| Lattice parameters | |
| $a$ (Å) | 3.9820(17) |
| $b$ (Å) | 3.8986(19) |
| $c$ (Å) | 10.666(6) |
| $α, β, γ$ (°) | 90, 90.240(18), 90 |
| $R1$ | 0.0455 |
| Completeness | 99.1% |
| Redundancy | 3.11 |
| $wR2$ | 0.1128 |
| Independent reflections | 341 |

| Atom | $x/a$ | $y/b$ | $z/c$ | Occupancy | $U_{eq}$(Å$^2$) |
|---|---|---|---|---|---|
| Eu | 0.754(3) | 0.75 | 0.7716(10) | 0.87(6) | 0.0151(12) |
| La | 0.72(2) | 0.75 | 0.760(6) | 0.13(6) | 0.025(15) |
| As1 | 0.2504(5) | 0.25 | 0.6334(2) | 1 | 0.0167(8) |
| As2 | 0.7502(6) | 0.25 | 0.9965(2) | 1 | 0.0252(8) |
| Fe | 0.2500(8) | 0.75 | 0.5000(3) | 1 | 0.0174(9) |

**Figure captions:**

**Figure 1.** (a) The optical image for the Eu$_{0.9}$La$_{0.1}$FeAs$_2$ single crystal. (b) The crystal structure for the EuFeAs$_2$ compound. (c) Powder X-ray diffraction patterns for the Eu$_{1-x}$La$_x$FeAs$_2$ polycrystalline samples.

**Figure 2.** Temperature dependence of normalized resistivity for the Eu$_{1-x}$La$_x$FeAs$_2$ polycrystalline samples.

**Figure 3.** Temperature dependence of magnetic susceptibility for the Eu$_{1-x}$La$_x$FeAs$_2$ polycrystalline samples.

**Figure 4.** Temperature dependence of magnetic susceptibility for the Eu$_{0.9}$La$_{0.1}$FeAs$_2$ single crystal.

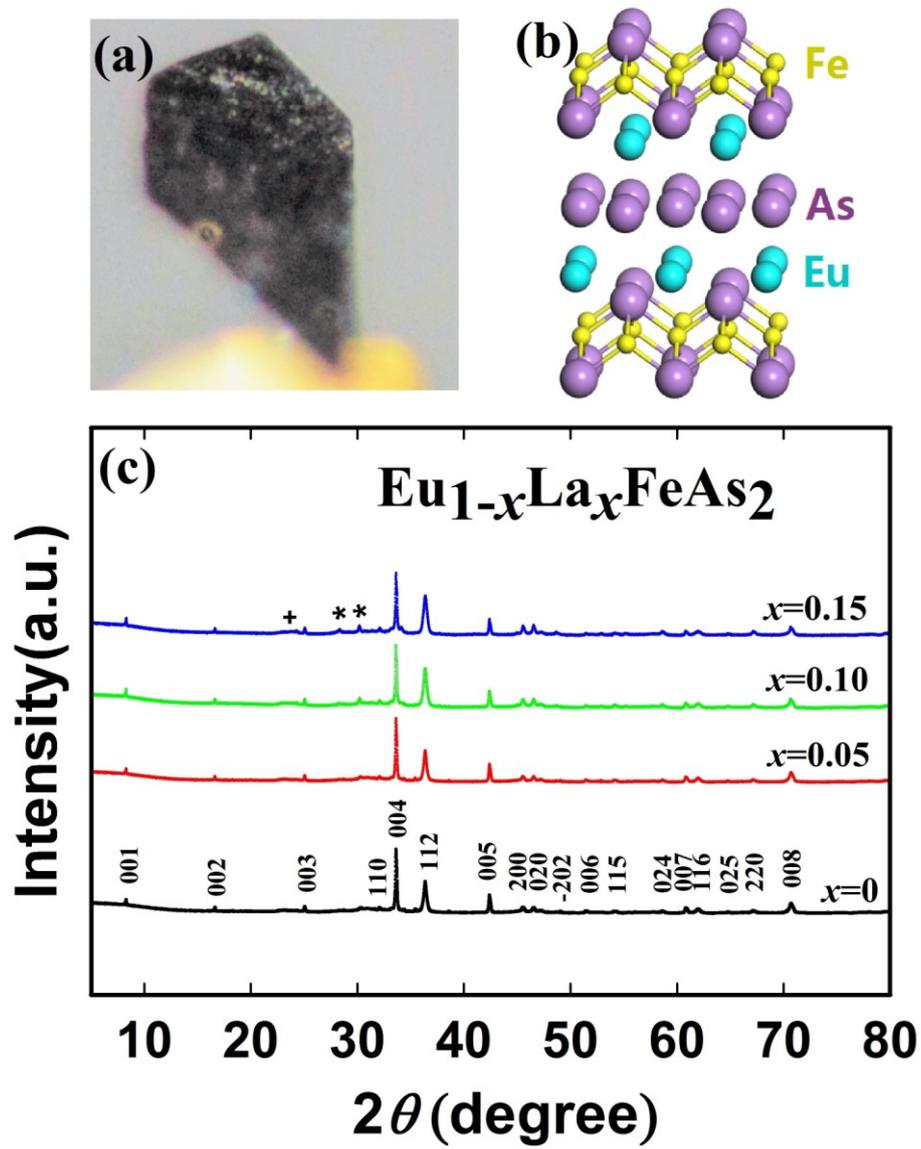

Fig. 1

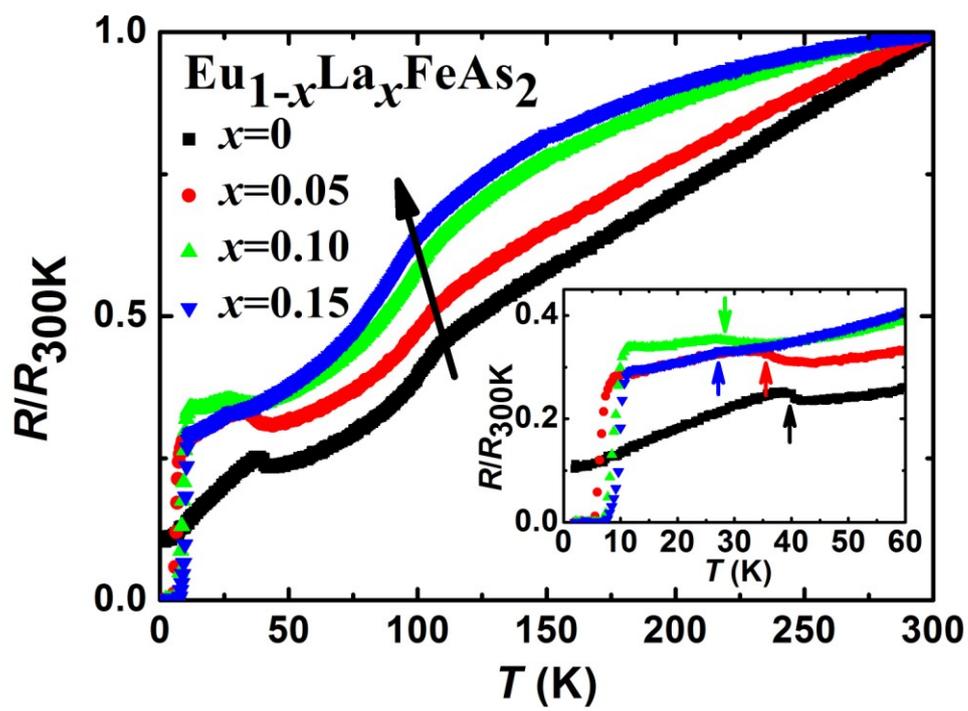

Fig. 2

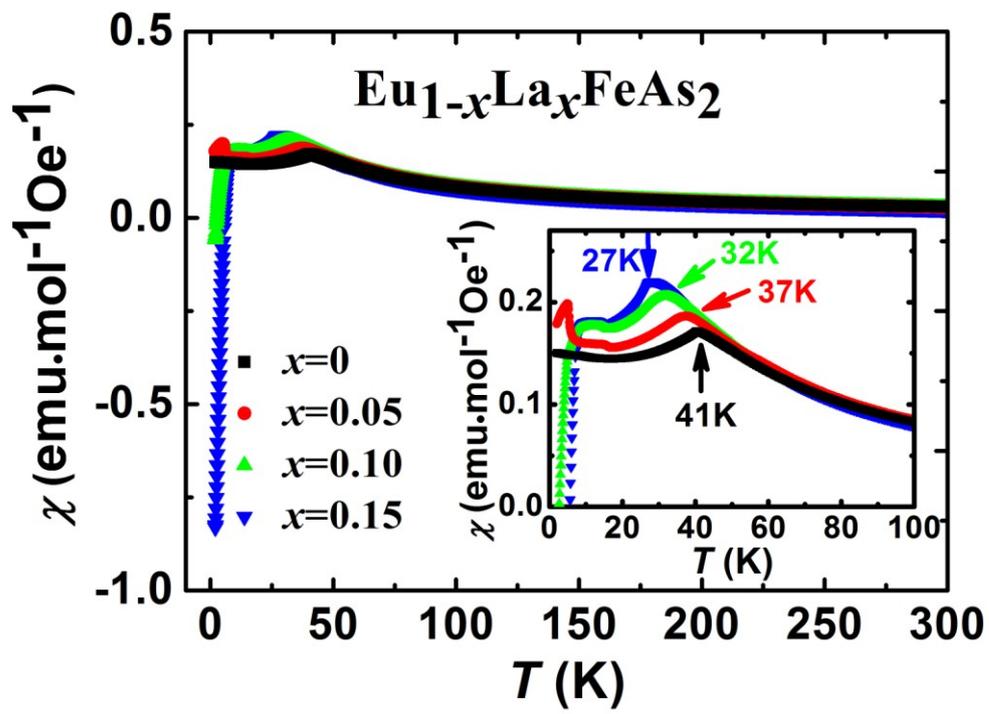

Fig. 3

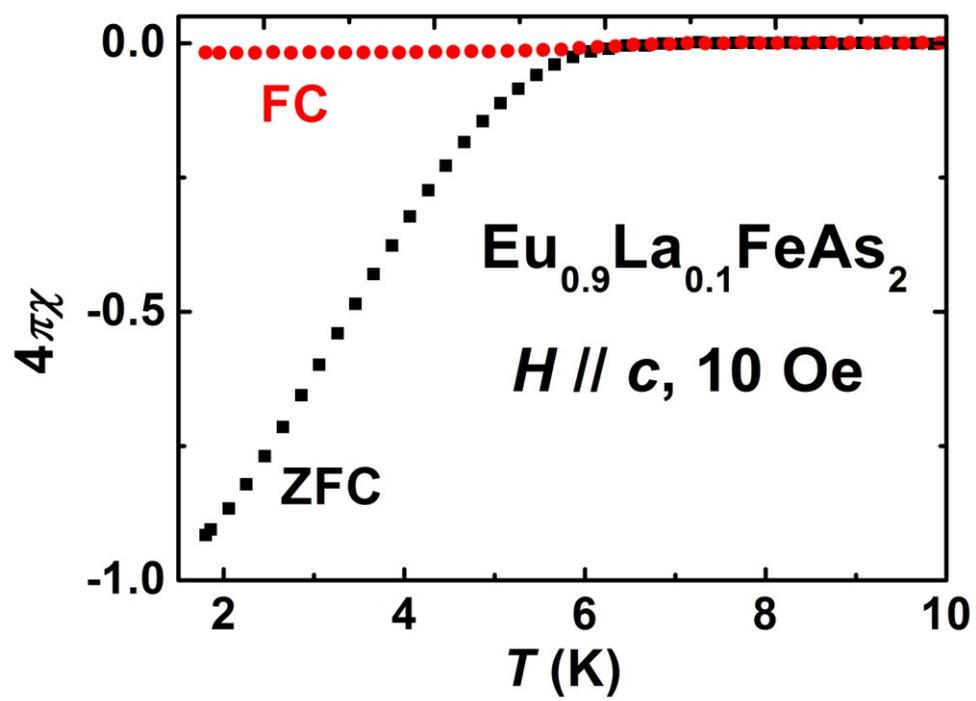

Fig. 4